\documentclass[showkeys,showpacs,twocolumn,amsmath,amssymb,epsf,aps]{revtex4}
\usepackage{dcolumn}
\usepackage{epsfig}
\usepackage{ulem}

\usepackage{graphicx}
\usepackage{longtable}
\usepackage{dcolumn}
\usepackage{bm}
\usepackage[usenames,dvipsnames]{xcolor}
\usepackage{setspace}

\begin{document}

\title{A Multiferroic Molecular Magnetic Qubit} 

\author{Alexander I. Johnson,$^{a}$ Fhokrul Islam,$^{b}$, Carlo M. Canali$^{b}$ and Mark R. Pederson$^{c}$}

\affiliation{$^{a}$~Department of Physics, Central Michigan University, Mt. Pleasant, MI 48859}                              
\affiliation{$^{b}$~Department of Physics and Electrical Engineering, Linneaus University, Kalmar, Sweden }
\affiliation{$^{c}$~Department of Physics, University of Texas El Paso, El Paso, TX 79968}                             

\date{\today}

\pacs{31.15.Ar, 31.70.Hq, 34.70.+e, 84.60.Jt, 87.15.Mi}

\keywords{Molecular Magnets, Density-Functional Methods}

\begin{abstract}

The chiral $Fe_3O(NC_5H_5)_3(O_2CC_6H_5)_6$ molecular cation, with C$_3$ symmetry, is composed of 
three six-fold coordinated spin-carrying Fe$^{3+}$ cations that
form a perfect equilateral triangle. Experimental reports demonstrating the 
spin-electric effect in this system also identify the presence of a magnetic uni-axis and
suggest that this molecule may be a good candidate for an
externally controllable molecular qubit. Here
we demonstrate, using standard density-functional methods, 
that the spin-electric behavior of this molecule could be even more interesting
as there are energetically competitive reference states associated with both 
high and low local spins (S=5/2 vs. S=1/2) on the Fe$^{3+}$ ions.
Each of these structures allow for spin-electric ground
states. {\color{black} We find that qualitative differences in the broadening of the Fe(2s) and O(1s) 
core levels, shifts in the core-level energies, and the magnetic signatures of the 
single-spin anisotropy Hamiltonian may be used to confirm whether a 
transition between a  high-spin manifold and a low spin manifold occurs.}

\end{abstract}
\maketitle

\section{Introduction and Motivation}

Molecules composed of spin carrying metal centers are ubiquitous in nature and play important roles in inorganic
chemistry and biophysics. Such systems are used to transport oxygen, convert water into oxygen and 
hydrogen~\cite{REF23,REF24,REF25,REF44}, and convert nitrogen to ammonia. 
The spin-spin and spin-orbit interactions associated 
with the heavier transition metal ions can provide non-destructive spectroscopic probes for understanding 
such chemical rearrangements~\cite{batool}. 
Further, the magnetic and spin-electric behaviors of these molecules suggest they are interesting candidates from the 
standpoint of quantum sensing, 
development of classical information storage systems, and also as possible qubits 
for quantum information applications~\cite{Komski,Leuen,qubit,Plass1}. 
While either integer or half-integer systems could be relevant to classical information 
storage systems, an analysis based upon a single-J Heisenberg Hamiltonian identifies triangular transition-metal 
complexes 

Two primary types of collective behaviors present themselves in molecules when inversion symmetry is absent. 
For molecules that contain metal
ions with large moments (more than two unpaired electrons per site) and relatively strong exchange-coupling, the
entire molecule behaves as a single spin at temperatures that  are small compared to inter-ionic exchange 
coupling.  If such molecules have uniaxial symmetries, spin-orbit coupling determines the 
magnitude and sign of D in the single-spin Hamiltonian (H$_A=DS_z^2$).~\cite{REF14,REF15,REF20,REF21,REF16,REF17,REF57,REF59}
If D is negative, the system exhibits easy axis, e.g. quasiclassical,  magnetic behavior. If D is positive, 
the system exhibits easy-plane magnetic behavior. If easy-axis magnetic behavior is observed
in these systems, the molecule is often referred to as a Type-1 molecular magnet or as an anisotropic molecular
magnet. 
The quantum states of such molecules can be switched through the
application of magnetic fields that are integer multiples of the anisotropy parameter (D).~\cite{REF20,REF21}

\begin{figure}[h]
\centering
\includegraphics[height=6cm]{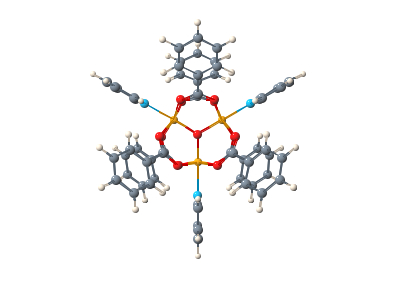}
\caption{
The $(Fe^{+3})_3O^{-2}(NC_5H_5)_3(O_2CC_6H_5^{-1})_6$ molecular cation  is shown above. 
Three Fe ions form a perfect equalateral triangle with an 
oxygen anion in the center of the molecule. Each of three equivalent Fe ions is coordinated by the central oxygen anion, 
four oxygen anions associated with two four different but symmetrically equivalent carboxylate groups (OOC-C$_6$H$_5$) and
one $NC_6H_5$ ligand.}
\label{fig1}
\end{figure}
Until recently, and in contrast to Type-1 molecular magnets, Type-2 molecular magnets or 
spin systems have been primarily found in
complexes containing an odd number of  metal ions that have small moments 
(one unpaired electron per site)~\cite{Dalal,V15,Loss,REF58,REF56,Boudalis1,Ardavan1,Turek1}.
{\color{black}For this case}, higher-energy charge-transfer
states between pairs of metal centers, otherwise known as the upper Hubbard band, 
are also coupled to the ground state by the spin-orbit interaction. This type of spin-orbit
effect is generally  referred to as the Dzylysinskii-Moriya (DM) interaction~\cite{Dzyal,Moriya} or 
antisymmetric exchange interaction. 
The occurrence of a lack of inversion
symmetry, and frustrated spin ordering provides a very different type of emergent behavior than what 
is observed in Type-1 molecular magnets.  Moreover, because there are multiple low-energy degenerate 
spin-ordered configurations, each of which lead to a net dipole moment, these systems exhibit the spin-electric
effect. 
{\color{black} The spin-electric effect, introduced in Refs.~\cite{Loss,Komski}  
arises when the spin-ordering in an equilateral triangle changes 
from having three ferromagnetically ordered spins
to one of the frustrated configurations with two parallel and one anti-parallel spin on their respective ions. 
For the remainder of this paper we refer to the non-ferromagnetically ordered configurations 
as either antiferromagnetic (AF) or frustrated.}
{\color{black} Shortly after the paper by Triff {\em et al}~\cite{Loss}, 
Islam {\em et al} quantified the behavior of these systems using 
the NRLMOL suite of codes and calculated the DM interaction using density-functional theory.~\cite{REF56,REF58}
This interaction is eventually reduced to $H_{DM}=\Delta_{Spin-Orbit} C_z S_z$ where C$_z$ is the chirality
operator and $\Delta_{Spin-Orbit}$ is a constant that depends on the 
spin-orbit interaction between the upper and lower Hubbard bands 
in the half-filled three-site C$_3$ system.  The result is that the chiral and antichiral states 
with the same M$_s$ are split by 2$\Delta_{Spin-Orbit}$.  
For a sufficiently large spin-orbit interaction this leads to a ground-state doublet 
composed of a chiral M$_s=\frac{1}{2}$ and anti-chiral M$_s=\frac{-1}{2}$ that can behave as a qubit.}                                           


The experimental demonstration of the spin-electric effect has occurred only recently.   For example,
Plass {\em et al}~\cite{Plass1} and Boudalis {\em et al}~\cite{Boudalis1} 
have experimentally identified  Cu-based and Fe-based molecular spin
qubits in which the transition metal ions are found on a triangular lattice in molecules 
with perfect C$_3$ symmetry. These
papers reiterate the need to construct such a qubit from metal ions that contain 
half-integer electron spin (S$_{loc}=\frac{2K+1}{2}$)
since, as also outlined below, such systems lead to a ground state with a 
total spin of S$_{tot}=\frac{1}{2}$.  

To date, most Type-2 molecular magnets, such as the one recently identified by Plass {\em et al}~\cite{Plass1},  
have been composed of metal ions that tend to carry small moments (e.g. Cu and V). However, 
the Fe-based system~\cite{Boudalis1}, introduced by Boudalis {\em et al}, 
is interesting due 
to the observations  of the magneto-electric coupling in a qubit composed of high-spin (S$_{loc}=\frac{5}{2}$) 
metal centers.
By subjecting the molecule to an electric field that is parallel to the plane of the 
three iron atoms, they have demonstrated that
the ground state couples to the externally applied electric field and comment that 
their work is the first experimental confirmation of the 
spin-electric effect. Boudalis {\em et al} point out that there have not been previous theoretical 
attempts at describing the spin-electric effect in systems such as Fe$^{3+}$ triangular systems. 

In this paper we perform calculations on two energetically competitive spin manifolds of the molecule 
introduced by Boudalis {\em et al}.  In contrast to other molecular magnets that have exhibited 
either Type-1 or Type-2 behavior but not both, our calculated results suggest that this molecule can exhibit both 
Type-1 and Type-2 behavior. 
In addition, the occurrence of two different spin manifolds, composed of high and low half-integer
triangular spins, yields two different states that both exhibit spin-electric coupling. 
This behavior is due to nearly degenerate qualitatively different  electronic configurations 
on the Fe$^{3+}$ ions. To further quantify this result, we perform calculations to determine 
the splitting between the two sets of Kramer doublets.

\begin{table*}[th]
  \caption{\ 
Atomic positions, local charges and local moments for the inequilvalent atoms for the low-moment and high-moment configurations of
the molecule.  The number of decimal places included in the table is technically necessary, with respect to round-off error, 
if one wishes to use all 12 symmetry operations that this molecule has. The local charges for the first nearest-neighbor 
oxygens, the nitrogens and the iron atom are very sensitive to the total spin as is the local charge on the Fe iron. 
{\color{black}The columns
presenting the local charges and local moments provide the integrated density ($\rho$) or spin density ($\rho_\uparrow-\rho_\downarrow$) 
within a sphere with a physically determined
radius around each atom and are provided to qualitatively show differences in the number 
of unpaired electrons and/or charge on each atomic site. 
The sphere radii, in Bohr, used for the table are 0.57, 1.46, 1.32 1.24 and 2.19 for H, C, O, N and Fe  respectively}.
 }
~ \\
~ \\
  \begin{tabular*}{\textwidth}{@{\extracolsep{\fill}}|c|c|c|c|c|c|c|}
\hline
Atom(${M_S=3/2}$)  &               X (Bohr)      &            Y (Bohr)         &             Z (Bohr)        &  Charge & Moment ($\mu_B$)  & Bond$_{Fe-X}$ (Bohr)   \\
\hline
Fe$_1$ &       -2.505257638898 &       0.000000000000 &       2.505257638898 &{\color{violet}{\bf  24.6076}}&{\color{red}{\bf -0.9059}}&6.14 \\
O$_1$ &       -0.000000000000 &      -0.000000000000 &       0.000000000000 &       5.8622&{\color{red}{\bf    -0.1589}}& 3.54\\
O$_2$ &        5.322330564106 &      -0.260403104702 &      -0.569772212054 &       5.9732&{\color{red}{\bf    -0.0081}}& 3.65\\
N$_1$ &       -5.469739941306 &       0.000000000000 &       5.469739941306 &       5.1263&{\color{red}{\bf     0.0089}}& 4.19\\
H$_1$ &       11.304895903521 &      -6.589849613853 &       2.357523144834 &      0.1674 &     0.0000 &\\
H$_2$ &        7.864556395661 &      -3.362449434801 &       2.251053480429 &      0.1688 &     0.0001 &\\
H$_3$ &       10.683035052801 &     -10.683035052801 &       0.000000000000 &      0.1680 &     0.0000 &\\
H$_4$ &       -6.543623208391 &      -6.620545602585 &       3.489866469741 &      0.1684 &    -0.0000 &\\
H$_5$ &       -8.368306728391 &     -10.916470039662 &       4.136202388450 &      0.1682 &     0.0000 &\\
H$_6$ &       14.719044124807 &       5.924135872740 &      -2.870772379330 &      0.1680 &    -0.0000 &\\
C$_1$ &        9.575468406012 &      -6.966039283004 &       1.304714561505 &      4.5269 &     0.0004 &\\
C$_2$ &        7.998708653694 &       3.103090515412 &      -1.792527622874 &      4.5186 &     0.0005 &\\
C$_3$ &       12.837466637187 &       5.150605294141 &      -2.536256048906 &      4.5148 &     0.0004 &\\
C$_4$ &        5.426718397270 &       1.954961030125 &      -1.516796337019 &      4.5631 &     0.0040 &\\
C$_5$ &       10.159365328460 &       1.761777636054 &      -1.028346613212 &      4.5154 &     0.0008 &\\
C$_6$ &       -6.513151601449 &     -10.695846403617 &       3.264169130844 &      4.5196 &    -0.0003 &\\
C$_7$ &        9.229044950899 &      -9.229044950899 &      -0.000000000000 &      4.5170 &     0.0007 &\\
C$_8$ &        7.662136522555 &      -5.149561540879 &       1.256287490837 &      4.5224 &     0.0009 &\\
\hline
Atom(${M_S=15/2}$) &          X            &        Y             &     Z                &  Charge    & Moment&Bond$_{Fe-X}$           \\
\hline
Fe$_1$ &       -2.624000490437 &       0.000000000000 &       2.624000490437 &{\color{violet}{\bf 24.1523}}&{\color{red}{\bf    -4.0536}}&6.42 \\
O$_1$ &       -0.000000000000 &      -0.000000000000 &      -0.000000000000 &      5.8806&{\color{red}{\bf    -0.4721}} &3.71\\
O$_2$ &        5.546433265240 &      -0.133871508300 &      -0.584097636058 &      5.9742&{\color{red}{\bf    -0.0924}} &3.88\\
N$_1$ &       -5.616413919617 &       0.000000000000 &       5.616413919617 &      5.1416&{\color{red}{\bf    -0.0480}} &4.23\\
H$_1$ &       11.439260476625 &      -6.717843055563 &       2.360708710532 &     0.1672 &    -0.0001 &\\
H$_2$ &        7.976303528027 &      -3.475887148864 &       2.250208189581 &     0.1684 &    -0.0004 &\\
H$_3$ &       10.802377433120 &     -10.802377433120 &      -0.000000000000 &     0.1673 &    -0.0000 &\\
H$_4$ &       -6.628367099071 &      -6.865310123732 &       3.552728696144 &     0.1681 &    -0.0000 &\\
H$_5$ &       -8.468314677454 &     -11.157230645789 &       4.201071032716 &     0.1682 &    -0.0000 &\\
H$_6$ &       14.965631738810 &       6.046722331729 &      -2.872187075351 &     0.1682 &     0.0000 &\\
C$_1$ &        9.705086882387 &      -7.083371945982 &       1.310857468206 &     4.5285 &    -0.0027 &\\
C$_2$ &        8.260207170085 &       3.208484962171 &      -1.843237245734 &     4.5238 &    -0.0118 &\\
C$_3$ &       13.086116921516 &       5.256980392042 &      -2.572156137428 &     4.5126 &    -0.0024 &\\
C$_4$ &        5.689882020246 &       2.067341855194 &      -1.555198309861 &     4.5576 &    -0.0099 &\\
C$_5$ &       10.415707071475 &       1.859209556094 &      -1.086370094156 &     4.5170 &    -0.0022 &\\
C$_6$ &       -6.616817839121 &     -10.946476994912 &       3.318646821096 &     4.5202 &    -0.0003 &\\
C$_7$ &        9.343716620569 &      -9.343716620569 &      -0.000000000000 &     4.5117 &    -0.0004 &\\
C$_8$ &        7.795420503774 &      -5.271264738051 &       1.262077882858 &     4.5261 &    -0.0023 &\\
\hline
\end{tabular*}
\label{table1}
\end{table*}

One of the ways that  high-spin and low-spin systems differ is that when spin-orbit is
turned on, there is indeed a difference between a low-moment C$_3$ system with one unpaired electron on each site and a high-moment C$_3$ system with five 
unpaired electrons on each site. The former problem maps onto a standard three-site one-band Heisenberg Hamiltonian and there is only one electron that 
can hop from one site to the next. However, for a high-moment system, there are several electrons on each site 
that can hop to other sites. In both high-moment and low-moment cases, these are high-energy charge transfer excitations. 
Further, for the high-spin case, it is possible, if not likely, that the energy to flip the spin of a single electron on a high-moment site is
the lowest electronic excitation in the problem. This is in contrast to the low-moment system where such a flip simply gives a different 
degenerate configuration. 
Alternatively, in the limit of a single
exchange-coupling parameter and no spin-orbit interaction it can be shown analytically that the lowest energy structure has 
a spin of $\frac{1}{2}$.

\section{Theory and Methodology }
The geometry of $(Fe^{+3})_3O^{-2}(NC_5H_5)_3(O_2CC_6H_5^{-1})_6$ cationic complex is shown in Fig.~\ref{fig1}. 
The point group symmetry of this molecule has twelve symmetry operations. When oriented so that the $<111>$ axis is along the 
cylindrical axis,  
the group may be generated from cyclic permutations, $(x,y,z)\rightarrow(z,x,y)$, 
an inverted non-cyclic permutation $(x,y,z)\rightarrow(-y,-x,-z)$ and the special rotation matrix, R$_Q$ below: \\
{\setstretch{1.5}
\[R_Q\equiv
\left(\begin{array}{rrr}
 \frac{1}{3}&-\frac{2}{3}&-\frac{2}{3} \\
-\frac{2}{3}& \frac{1}{3}&-\frac{2}{3} \\
-\frac{2}{3}&-\frac{2}{3}& \frac{1}{3} 
\end{array} \right) 
\]
}
%
{\color{black}
The matrix that we call $R_Q$ has interesting properties which we want to point out in the event there is deeper relevance. 
One third of the matrix elements have a magnitude of 1/3. Two thirds of the matrix elements have a 
magnitude of 2/3. The sum of each row is -1. The sum of each column is -1. The sum of the 
diagonal elements is 1. The matrix is the transpose of itself and is also its own inverse. 
The determinant of the matrix is -1 which is of course expected of a matrix-representation of a symmetry operation. The determinant of
any 2x2 minor matrix is either 1/3 or -2/3. The eigenvalues of this symmetric  matrix all have a magnitude of 1 with 1/3 of the 
eigenvalues being negative and 2/3 of the eigenvalues being positive.
}
The resulting group of order 12 leads to a total of six representations with degeneracies of (1,1,2,2,1,1). 
None of these representations are either even or odd under inversion. For an isolated $Fe^{+3}$ atom, 
the ground state of the system has an outer valence configuration of $3d^{\uparrow\uparrow\uparrow\uparrow\uparrow}$ 
leading to a high net spin of $S=5/2$.  However, a higher energy configuration of 
$3d^{\uparrow\uparrow\uparrow}3d^{\downarrow\downarrow}$ with a low net spin of $S=1/2$ is also electronically and 
magnetically stable but approximately 2-3 eV higher in energy than the high-spin $Fe^{+3}$ cation. Our results 
indicate that when ligated according to the figure, the relative stability of the high- and low-spin is 
changed significantly. In Table~\ref{table1} we present the local charges and moments for each of the inequivalent atoms
that we have found in calculations for the high- and low- moment manifolds.  We also compare the 
five Fe-O and Fe-N nearest-neighbor distances for the two structures.

\begin{table*}[th]
\caption{\ 
Magnetic and spin-electric properties for two nearly degenerate spin manifolds of
the cation. 
The first manifold  has a local spin and valence of 
$^{1/2}3d^{\uparrow\uparrow\uparrow\downarrow\downarrow}$ on each Fe ion. 
The second manifold has a local valence of 
$^{5/2}3d^{\uparrow\uparrow\uparrow\uparrow\uparrow}$ on each Fe ion. {\color{black} For both cases the energy splitting between the 
antiferromagnetic and ferromagnetic structure ($\Delta^{AF/FM}$) and the 
magnetic anisotropy energy (MAE) is given}. The spontaneous dipole moments acquired
by the antiferromagnetically ordered structures are qualitatively different in magnitude.
The calculated
local moments (within a sphere of radius 2.19 Bohr) are consistent with the local valences. 
In both cases, the dipole moments result from 
movement of electrons from the two ions with paired spins to the ion with an unpaired spin as in the case of 
Ref.~\cite{REF58}.  For calculations of
J, we use the convention that the Heisenberg Hamiltonian is given by 
$H=E_o+J( {\bf S_1} \cdot {\bf S_2}+ {\bf S_2} \cdot {\bf S_3}+ {\bf S_3} \cdot {\bf S_1})$. 
In other words both manifolds prefer antiferromagnetic ordering.
}\label{table2}
~ \\
~ \\
\begin{tabular}{|c|l|l|r|l|c|l|l|l|l|c|}
\hline                      
$S_{loc}$&E$_o$(eV) &J(eV)&MAE (K)  &$\Delta^{AF/FM}$&AF Dipole (au)& $\mu^{FM}$ &$\mu^{AF}_1$  & $\mu^{AF}_2$& $\mu^{AF}_3$&$E_o+J[\frac{3}{4}-3S_{loc}(S_{loc}+1)]$\\          
\hline                      
 1/2   & 0.000     &0.053& 3.8 (EA)&-0.053 & 0.091        & 0.91       & 0.97         &-0.93        &-0.93 &-0.08 (eV)\\
 5/2   & 0.095     &0.019&-8.5 (EP)&-0.424 & 0.139        & 4.05       & 3.94         &-3.84        &-3.84 &-0.39 (eV)\\
\hline
\end{tabular}
\end{table*}
The calculations presented here use approximations to density-functional theory (DFT)~\cite{REF26o,REF26,REF27}, 
at the generalized-gradient level, to describe the electronic and magnetic structure of the molecular cation. The 
NRLMOL computational code employs Gaussian orbitals to represent the Kohn-Sham orbitals.~\cite{REF28,REF29,REF30}
The basis sets that are used in NRLMOL are roughly triple to quadruple zeta quality~\cite{REF30}. 
The PBE-GGA density functional approximation is used for the exchange correlation functional in these
self-consistent-field (SCF) calculations.  
For the iron atom, we have used 20 bare Gaussians with exponents varying from $0.12050$ to $0.38667\times10^7$ to
construct a contracted basis set composed of 7 s-type, 5 p-type , and 4 d-type contracted orbitals.
For simplicity in describing the basis sets on other atoms we refer to this as
[Fe(20): $0.12050 - 0.38667\times10^7$, 7s5p4d].
With this notation the oxygen, nitrogen, carbon, and hydrogen basis sets are designated as:
[O(13): $0.10492 - 0.61210\times10^{5}$, 5s4p3d],
[N(13): $0.09411 - 0.51751\times10^{5}$, 5s4p3d],
[C(12): $0.07721 - 0.22213\times10^5$, 5s4p3d] and [H(6): $0.22838-0.77840\times10^2$, 4s3p1d] respectively.
The Limited-Memory Broyden-Fletcher-Goldfarb-Shanno (LBFGS) scheme programmed by Liu and Nocedal~\cite{REF31} has been used for the optimization of
structure with energy and force convergence criteria of $0.10000\times10^{-5}$ Hartree
and .003 Hartree/Bohr for consistency.

Spin polarized calculations have been performed with the inclusion of spin-orbit coupling for the evaluation of magnetic
properties~\cite{REF15}. 
\section{Results and Analysis}
{\color{black} 
Generally speaking, to determine the relevant parameters for the spin Hamiltonians for a triangular lattice of identical spins ($S_{loc}$=1/2 or 5/2 here), 
one calculates the energy of the
ferromagnetically ordered state $E^{FM}$ and the anti-ferromagnetically ordered state $E^{AF}$. Then based on the assumption of a
Heisenberg Hamiltonian of the form 
$H=E_o+J[{\bf S_1} \cdot {\bf S_2}+ {\bf S_2} \cdot {\bf S_3}+ {\bf S_3} \cdot {\bf S_1}]$, one can determine that 
$E_o=(E^{FM}+3E^{AF})/4$ and $J=(E^{FM}-E^{AF})/(4S_{loc}^2$).
For our calculations, we find energies of 
$E^{FM}(S_{loc}=1/2)=-7129.712342$ and $E^{FM}(S_{loc}=5/2)=-7129.697517$ Hartree. 
For each of these cases we then calculate
the antiferromagnetically ordered structures finding energies of 
$E^{AF}(S_{loc}=1/2)=-7129.714280$ and $E^{AF}(S_{loc}=5/2)=-7129.714586$ Hartree. 
After extracting $E_o$ and J for both the high-spin and low-spin configurations  
we can arbitrarily shift
both spin Hamiltonians by a constant so that $E_o(S_{loc}=1/2)\equiv 0$.  The resulting values of J and $E_o$ are given in Table~\ref{table2}.
This Heisenberg Hamiltonian may be more simply represented according to:
\begin{eqnarray}
H=E_o(S_{loc})+\frac{J}{2}\{[\Sigma_i {\bf S_i}]^2-\Sigma_i S_i^2\}.
\end{eqnarray}
The resulting eigenstates of the  above Hamiltonian, arising from a specific $S_{loc}$, 
are also eigenstates of $S_{tot}^2$, ${S^z_{tot}}$. In the absence of
anisotropy, the energies are given by
\begin{equation}
E=E_o(S_{loc})+\frac{J}{2}[S_{tot}(S_{tot}+1) - 3 S_{loc}(S_{loc}+1)]
\end{equation}
This affirms  that regardless of $S_{loc}$ the ground-state energy corresponds to $S_{tot}=1/2$. 
This equation allows us to determine the relative energies of the two sets of Kramer doublets 
that are associated with the two different spin manifolds. These are presented in the last 
column of Table~\ref{table2}.
}
{\color{black}
In Table~\ref{table2} we present results for the energies of the high-spin and low-spin manifolds as a 
function of spin orderings.  
Also presented in Table~\ref{table2} are the magnetic anisotropy energy (MAE) for each 
spin manifold and the energy gap ($\Delta^{AF/FM})$) between ferromagnetic and antiferromagnetic structures.
To definitively affirm our 
local-spin assignments of S=1/2 and S=5/2 metal-centers and the spin orderings for various 
calculations, we also present the local moments, labeled $\mu_i^{AF}$ or $\mu_i^{FM}$, in Table II.  
Confirmation of these assignments are necessary to justify the form of the Heisenberg Hamiltonian 
we have utilized in this analysis.  While the total number (spin moment) of unpaired electrons 
(3 for the low-spin manifold and 15 for the high-spin manifold) are well defined quantities 
and indicative of spin 1/2 and 5/2 metal centers, this additional information is required to unambiguously 
demonstrate, computationally, that the electrons participating in spin polarization are indeed 
primarily associated with the metal ions. Our integrated net spins in non-space filling spheres 
(R=2.19 au) are in the range of $\mu = 3.84-4.04$ and $\mu=0.91-0.97$ for the high-spin and 
low-spin systems. The lack of local moments on other sites (See Table~\ref{table1}) confirms that the Heisenberg-model 
employed here is sufficient for the analysis. 

As shown in Table~\ref{table2}, the dipole moment for 
the antiferromagnetic high-spin (5/2) manifold is larger by a factor of 1.5.  In previous papers~\cite{Loss,REF58,REF56} 
it has been noted that the dipole moment of the generally non-stationary antiferromagnetic 
state is the key indicator of a good spin-electric system composed of low-spin metal centers. 
We emphasize that these antiferromagnetic states are only eigenstates of the system in the 
high-field limit. In addition to differing spin-electric coupling strengths in the low-field limit, 
this result suggests that the system could be driven from the low-spin to high-spin manifold by 
application of an electric field. As such this system also presents the possibility for a uni-directional electrostic spin crossover effect that has been discussed earlier in applications to explicitly polar magnetic molecules~\cite{Baadji}. }

\begin{table}
  \caption{\ Energy (eV) of ferromagnetically ordered molecule as a function of $M_S$ and the geometry. This molecule shows spin 
crossover behavior as a function of the geometry which suggests that the ground magnetic state will change when pressure is applied and 
that it could be sensitive to size of the charge-compensating counter anion in the  unit cell.
 } \label{table3}
~ \\
~ \\
\begin{tabular}{|l|l||l|l|l|}
\hline
$M_S$&  $\Delta E^{3/2}$ &$\mu^{3/2}$               & $\Delta E^{15/2}$&$\mu^{15/2}$\\
\hline
 3/2 &  0.000  &0.91      & 1.602  &0.93   \\
 6/2 &  0.891  &1.88      & 1.792  &1.95  \\
 9/2 &  1.277  &2.73      & 1.226  &2.82  \\
12/2 &  1.865  &3.45      & 0.924  &3.51  \\
15/2 &  1.930  &4.05      & 0.404  &4.05  \\
\hline
\end{tabular}
\end{table}

There are in fact other spin-ordered solutions, with qualitatively different moments, than discussed in detail here, that we have found
during the course of these calculations. Because of this we have performed fixed moment ferromagnetic single-point calculations, as a function of total moment,
at the low-spin and high-spin equilibrium geometries. In Table~\ref{table3} we present the energies as a function of total moment. The results show that
it is indeed only the lowest-spin and highest-spin configurations that are expected to be energetically competitive ground states.

From Table~\ref{table1} we observe that a spin moment appears on the central atoms for the high-spin case but not for the low-spin case. This moment has the same sign as the
moment on the Fe ions which could simply mean that some of the valence charge on the Fe ions is leaking onto the central oxygen. If this is the case, it could mean that there is
a greater degree of Fe $4s$-$3d$ hybridization for the case of the high-spin manifold.  To make links with future experiments,  we provide predictions of the relative energies
of the Fe $1s$ and $2s$ states and the O $1s$ states as a function of spin manifold.  We find that the Fe 
1s-core levels for the high spin manifolds shift downward by 1.4 eV. Similarily the
Fe $2s$ core levels move downward by approximately 1.4 eV. However, in contrast to  the Fe 1s-core levels, 
the 2s-core levels exhibit a qualitatively different broadening for the two spin manifolds. 
For the low-spin manifold the width of the Fe $2s$ core energies is approximately 0.5 eV. But for the high-spin manifold, the broadening increased to 2.2 eV. The broadening is consistent 
with spin polarization rather than chemical bonding. 
The oxygen 1s states also show a larger broadening (0.82 vs 0.16 eV) for the case of the high-spin manifold 
and are shifted upward, rather than downward, by about 0.47 eV relative to the low-spin manifold. 
A perfect ionic model would predict that the 1s oxygen levels would shift by 1.1 eV if we only
consider the change in Coulomb potential due to a +3 Fe cation at the origin.
{\color{black} By broadening, we are referring to changes in energies of various core states 
due to a combination of exchange coupling and geometrical changes in the molecules.  } 
\section{Summary}
To summarize, the Fe$^{III}$ ions, arranged at the vertices of an equilateral triangle, in the cationic molecule studied here can exist with valences of 
3d$^{ \uparrow \uparrow \uparrow \uparrow \uparrow}$ or 3d$^{ \uparrow \uparrow \uparrow \downarrow \downarrow}$. Our geometries, optimized for each spin 
configuration, find that the lowest energy Kramer doublet states for each system are close in energy                                                       
but with the high-moment case slightly lower in energy (~0.3 eV). {\color{black} Both of these degenerate Kramer doublet states can be manipulated with an electric field.} 
{\color{black} As discussed in Refs.~\cite{V15,REF58,REF56,ERuiz,KPark}}, it is generally expected that density-functional methods, without corrections for the onsite Coulomb repulsion, 
will overestimate exchange-coupling parameters by a factor of 2-3 which would then adjust the energy splitting to a number that is close to room temperature. 
There are other factors that usually do not enter in to determining the optimal ground-state reference energy that could then play a role in fully determining
the splitting between the two sets of Kramer doublets that arise from the two different reference states. Other factors that should be considered include 
zero-point vibrational energies and average spin-orbit energies. The expectation is that vibrational effects would slightly stabilize the low-spin manifold and
that inclusion of the total spin-orbit energy would slightly favor the high-spin manifold. Work on quantifying this result
is in progress.

Regardless of which reference state is lower in energy, the presence of nearly degenerate reference states, composed of high-spin and low-spin centers, 
is of potential interest for quantum sensing applications. The low-spin case offers the 
possibility of a molecular magnet that has weak easy-axis anisotropy and that would therefore be switchable through the applications of magnetic fields as
well as electric fields.  The low-spin and high-spin case offers the possibility of two different spin-electric signatures.
Prospects for practically switching between these two states appear to be good. The overall size of the molecular cation 
depends on the moments which means that the 
Madelung stabilization of
the low-spin molecule, in a crystal, 
could be greater if counter anions of differing size can be used. 
In addition to more experiments on this interesting molecular magnetic qubit, 
it will be important to consider the use of higher-level quantum chemistry methods,  
and improved density-functional based methods, such as those obtained from the 
Fermi-L\"owdin-Orbital-Self-Interaction-Corrected 
(FLOSIC)~\cite{REF33,REF34,REF35,REF36,REF37,REF40,REF41,REF41b}.
Finally the 
inclusion of non-collinear mean-field methods, with comparison to multiconfigurational methods, will help to realize 
quantitative computational understanding of this qubit. Due to the importance of rigorously 
understanding decohering mechanisms in these systems,
inclusion of group-theoretical symmetrization methods have been used here. Fully incorporating such techniques into non-collinear and multi-configurational approaches 
should enhance the effectiveness for applications to quantum sensors and qubits.
\section{Acknowledgements}
{\color{black}
A.I. Johnson thanks the Michigan-State Computer Center for computer time used on this project.
M.R.P. was supported from startup funds provided by the University of Texas-El Paso and the Texas Science and Technology
Aquisition and Retention (STARS) Program.
CMC and FI are supported by the Faculty of Technology at Linnaeus University, the Swedish Research Council (VR) through Grant No. 621-2014-4785, and by the Carl
Tryggers Stiftelse through Grant No. CTS 14:178. Computational resources for preliminary calculations have been provided by the Lunarc Center for Scientific and 
Technical Computing at Lund University.}

\end{document}